\documentclass[10pt,twoside]{amsart}
\advance\oddsidemargin by -1.0cm
\advance\evensidemargin by -1.0cm
\advance\topmargin by -1.0cm 
\usepackage{amssymb}
\usepackage{amscd}
\usepackage{array}
\usepackage{epsfig}
\let\mathg\mathfrak
\theoremstyle{plain}
\newtheorem{cor}{Corollary}
\newtheorem{lem}{Lemma}
\newtheorem{thm}{Theorem}
\theoremstyle{remark}

\newtheorem{rem}{Remark}
\newcommand{\bdm}{\begin{displaymath}}
\newcommand{\edm}{\end{displaymath}}
\newcommand{\ba}[1]{\begin{array}{#1}}
\newcommand{\ea}{\end{array}}
\newcommand{\bea}[1][]{\begin{eqnarray#1}}
\newcommand{\eea}[1][]{\end{eqnarray#1}}

\newcommand{\bb}{\begin{bundle}}
\newcommand{\eb}{\end{bundle}}
\newcommand{\bcen}{\begin{center}}
\newcommand{\ecen}{\end{center}}
\newcommand{\btab}{\begin{tabular}}
\newcommand{\etab}{\end{tabular}}

\newcommand{\x}{\times}
\newcommand{\op}{\oplus}
\newcommand{\ox}{\otimes}

\newcommand{\ra}{\rightarrow}
\newcommand{\lra}{\longrightarrow}

\newcommand{\Lra}{\Leftrightarrow}
\newcommand{\lmapsto}{\longmapsto}

\newcommand{\rx}{\rtimes}

\newcommand{\qqs}{\forall}
\newcommand{\lan}{\left\langle}
\newcommand{\ran}{\right\rangle}
\newcommand{\tssum}{\textstyle\sum}
\newcommand{\C}{\ensuremath{\mathbf{C}}}
\newcommand{\R}{\ensuremath{\mathbf{R}}}

\newcommand{\N}{\ensuremath{\mathbf{N}}}
\newcommand{\Z}{\ensuremath{\mathbf{Z}}}
\newcommand{\Cycl}{\ensuremath{\mathcal{C}}}
\newcommand{\mc}[1]{\mathcal{#1}}
\newcommand{\mrm}[1]{\mathrm{#1}}

\newcommand{\vphi}{\ensuremath{\varphi}}    
\newcommand{\vrho}{\ensuremath{\varrho}}
\newcommand{\lalg}[1][g]{\ensuremath{\mathg{#1}}}

\newcommand{\kom}{\ensuremath{\mathcal{K}}}       
\newcommand{\cent}[1][G]{\ensuremath 
 {\mathcal{Z}(#1)}}
\newcommand{\cento}{\ensuremath      
 {\mathcal{Z}_0}}\newcommand{\udef}[1][n]{\ensuremath 
 {\mathord{\mathg{u}(#1)}}}
\newcommand{\Udef}[1][n]{\ensuremath                    
 {\mathord{\mathrm{U}(#1)}}}
\newcommand{\sudef}[1][n]{\ensuremath                   
 {\mathord{\mathg{su}(#1)}}}
\newcommand{\SUdef}[1][n]{\ensuremath                   
 {\mathord{\mathrm{SU}(#1)}}}

\newcommand{\symp}[2][n]{\ensuremath                    
 {\mathsf{S}^{#1}#2}}

\newcommand{\extp}[2][n]{\ensuremath                    
 {{\textstyle\bigwedge\nolimits^{#1}} #2}}

\newcommand{\SOdef}[1][n]{\ensuremath
 {\mathord{\mathrm{SO}(#1)}}}
\newcommand{\grpSM}{\ensuremath                         
 {G_{\mathit{SM}}}}
\newcommand{\lalgSM}{\ensuremath                        
 {\mathg{g}_{\mathit{SM}}}}
\begin{document}
\setcounter{equation}{0}
%
%
\thispagestyle{empty}
%

\hbox to \hsize{%
  \vtop{} \hfill
  \vtop{\hbox{MPI-PhT/97-44}}}
\title[Covering groups and the standard elementary particle model]{
Covering groups of the gauge group for the standard
elementary particle model}
%
%
%
\dedicatory{Ilka Agricola (\texttt{agricola@mathematik.hu-berlin.de}),\\
Math. Institut der Humboldt-Universit\"at zu Berlin,\\
D-10099 Berlin, Germany}
%
\thanks{This paper summarizes some of the results of my Diploma
Thesis, supervised by Heinrich Saller and
presented at the Dept.\ of physics, 
Ludwig-Maximilians-Universit\"at M\"unchen, Germany, 1996.}
\thanks{On leave of absence from Max-Planck-Institut f\"ur Physik,
F\"ohringer Ring 6, D--80805 M\"unchen, Germany}
\begin{abstract}
We determine all Lie groups compatible with the
gauge structure of the Standard Elementary Particle Model (SM) and
their representations. The groups are specified by congruence
equations of quantum numbers. By
comparison with the experimental results, we single out one Lie group \grpSM\
and show that this choice implies certain old and new correlations between the
quantum numbers of the SM quantum fields as well as some hitherto unknown
group theoretical properties of the Higgs mechanism.
\end{abstract}
\maketitle
%
\pagestyle{headings}
%
%
%
\section{Introduction}\label{SM}
\subsection{The Lie algebra of the Standard Model}
The usual and well known choice for the Lie algebra of internal symmetries 
within the Standard Elementary Particle Model (SM) is
 \begin{equation}\label{SM-lalg}
 \lalg_{SM} \ = \ \udef[1]_Y \ \op \ \sudef[2]_T \ \op \ \sudef[3]_C\, ,
 \end{equation}
where we denoted by $Y$, $T$ and $C$ the internal symmetries hypercharge,
isospin and colour respectively. For many purposes, the knowledge of the
\emph{local} structure of the theory, that is the Lie algebra $\lalg_{SM}$
is sufficient. For example, the dynamics of any
quantum field theory (Feynman rules) depend only on them. However,
some investigations of the gauge structure of a QFT require knowledge
of the \emph{global} structure, a fact we are well accustomed with from
the manifold side. Since a
standard result of Lie theory states that a Lie group is \emph{not}
globally determined by its Lie algebra, the
main idea of this paper is to find
where in the gauge structure of the SM the Lie \emph{group} of internal
symmetries appears, which of the possible choices for it is the
``true'' Lie group $G_{SM}$ of the standard model, and what further conclusions
this choice implies. Since we always want particle multiplets to be
 finite dimensional, we restrict our attention  to \emph{compact} Lie groups.

\subsection{Results}
After compiling the necessary mathematical definitions and
prerequisites in Section \ref{notations}, we give a complete
classification of all compact Lie groups compatible with the internal
gauge structure of the SM (Thm.~\ref{abstrclass}) and a practical
description of their representations via  congruence equations between 
quantum numbers (``integrability conditions'', Lemma \ref{intcond}), thus
generalizing and correcting some former results by L.~O'Raifeartaigh
\cite{O'Raif79},\cite{O'Raif} and  J.~Hucks \cite{Hucks91}. 

By comparison with the well known particle content of the SM 
(Table \ref{partpropbis}), exactly
one of these Lie groups (called \grpSM) is singled out for a
``minimal'' description in Section \ref{minLiegrp}; we furthermore 
show that its representation ring is generated by
these particles together with their antiparticles. Assuming from there on
that \grpSM\ is indeed the Lie group of the Standard Model, we can then
show that
confined quark states have necessarily integer electric charge
(Lemma~\ref{quarkconfinlem}). Futhermore, this choice implies that 
although the hypercharge $y$ of a particle may be fractional,
its product with the dimensions of the isospin and colour
representations is always an integer. Equivalently, the sum over all electric
charges inside any particle multiplet is necessarily integral
(Thm.~\ref{teilerdim}, Cor.~\ref{teilerdimSM}).

We then show that \grpSM\ is actually isomorphic to
$\mrm{S}(\Udef[2]\x\Udef[3])$ 
(Section~\ref{addreal}) as well as to the Kronecker product of $\SUdef[2]$ and
$\SUdef[3]$ with canonical scalar $\Udef[1]$ action (Section~\ref{multreal}),
thus allowing a second proof of Theorem \ref{teilerdim} in
Remark~\ref{secondproof}. We close with the surprising fact that 
for electroweak $\Udef[2]$, the usual definition for electric
charge yields a complement of $\SUdef[2]_T$, whereas hypercharge does not.
This decomposition is equivalent to the Gell-Mann-Nishijima
Formula (Section~\ref{zerlegkompgrp}).

\section{Notations and mathematical prerequisites}\label{notations}
\subsection{Cyclic groups, tori and hypercharge}
%
For finite cyclic groups, we will always use the notation
 \begin{equation}\label{centun}
 \Cycl_n\ = \ \left\{\zeta^m_n \ \bigm|\ \ m=1, \ldots,n \right\}, \quad
 \zeta_n :=\exp{\frac{2\pi i}{n}}\, .
 \end{equation}
 The irreducible unitary characters of $\Udef[1]$ are given by a
 discrete quantity $\lambda$ (sometimes called \emph{winding number})
 \begin{equation}\label{udefonechar}
 \chi_{\lambda}(e^{i\theta})\ =\ e^{i\lambda\theta},\ \lambda\in\Z 
 \end{equation}
 where we will succinctly write the corresponding $\Udef[1]$-module as
 $\C_{\lambda}$. 
 For historical reasons, physicists label $\Udef[1]_Y$-representations 
 not by the integer $\lambda$, but rather by ``hypercharge'' $y=\lambda/6$.
 The universal covering group \R\ of $\Udef[1]$ has the smooth
 unitary irreducible representations $x \mapsto e^{ix\Lambda}, \ \Lambda\in\R$,
 which can only give characters of the factor group $\R/\mu\Z$, 
 $\mu\in\R^{\x}$, when the condition
 $\mu\Lambda \in 2\pi\Z$
 holds. Thus the mere transition from the noncompact real line \R\ to
 the compact group $\Udef[1]$ accounts already for the discreteness of
 the corresponding labeling parameter for its representations
 (compare this with \cite{Hucks91}).

\subsection{Irreducible representations of $\mathfrak{su}(n)$}\label{irredsun}
%
 We shall describe any irreducible 
 unitary (necessarily finite dimensional) representation $\vrho$ of $\sudef$
 by its highest weight, written as a linear combination with non-negative
 integral coefficients of some choice of fundamental weights, and refer to 
 these coefficients as the \emph{Dynkin indices} of the representation
 $\vrho$.
 If we denote by $V$ the $n$-dimensional standard representation of
 $\sudef$, we know that the representation $\vrho$ with Dynkin indices 
 $(a_1,\ldots,a_{n-1})$ will appear as a subrepresentation of the tensor 
 product
 \begin{equation}\label{tensorrep}
 T_{\vrho}\ =\ \symp[a_1]V\ox \symp[a_2](\extp[2]V)\ox \cdots \ox 
 \symp[a_{n-1}](\extp[n-1]V)
 \subset V^{\ox r},\ r= \tssum_{j=1}^{n-1} j a_j\, .
 \end{equation}
\begin{rem}\label{repsize}\textit{Size of a representation.}
 We will refer to the number $r$ as the \emph{size} of the representation 
 and denote its congruence class modulo $\dim\ V$ by $\bar{r}$.
 In terms of Young diagrams, $r$ corresponds to the number of squares
 in a diagram. Physically speaking, we will prove that the
 representations of the center of $G_{SM}$ will be
 completely determined by this number. Notice that the element 
 $\zeta_n\cdot 1_n$
 of \SUdef\ will act on the tensor representation $T_{\vrho}$ or any
 other subrepresentation of $V^{\ox r}$ as multiplication by
 $\zeta^{\bar{r}}_n$; in particular, this implies that 
 the trivial
 representation can only appear in $T_{\vrho}$ ofequation~\eqref{tensorrep}
 if $\bar{r}=0$ (the converse, of course, being false).
\end{rem}
\subsection{Irreducible isospin and colour representations}
The representations of the rank-1 Lie algebra $\sudef[2]_T$ can
be labeled by one Dynkin index; again for historical reasons,
physicists use instead half its Dynkin index $t\in\frac{1}{2}\N_{0}$. 
For $\sudef[3]_C$, the Cartan subalgebra is two dimensional 
and  one may choose the standard gluon fields $g$ and $\bar{b}$ as
fundamental weights 
with corresponding Dynkin indices $i$ and $j$, which are exactly the
colour charges listed in Table~\ref{partpropbis}. The sizes of any 
$\sudef[2]_T$- or $\sudef[3]_C$-representation will be denoted by
$r_T$ or $r_C$, respectively. We denote the electric charge in
multiples of $e$ be $q$ and will
will assume the Gell-Mann-Nishijima formula
$q=y+t_3$ to be valid throughout this article.

\begin{rem}\label{quarkconfinrem}\textit{Quark confinement.}
The postulate of quark confinement that \textit{any physical quark state
must be a color singlet} means that both colour charges $i$, $j$ have
to vanish. Since physical states appear in tensor representations like
$T_{\vrho}$ or any other subrepresentation of $V^{\ox r}$,
Remark \ref{repsize} implies that such an $\sudef[3]$
tensor representation can contain a colour singlet only if\linebreak
$r_C\equiv 0\bmod\, 3$. 
\end{rem}

\section{Determining all Lie groups with Lie algebra \lalgSM}
\subsection{Determining all (compact) Lie groups of a reductive Lie
algebra}\label{allLiegrpsabstract} 
%
The universal (connected) covering group $\tilde{G}$ of any reductive
\R-Lie algebra \lalg\ is the direct product of its commutator subgroup and
the connected component of the identity of its center:
 \bdm
 \tilde{G} \ \cong\ \tilde{G}_c \x \cent[\tilde{G}]_0,\ \cent[\tilde{G}]_0
 \cong\ \R^n.
 \edm
 For the full center, this means
 \bdm
 \cent[\tilde{G}] \ \cong\ \cent[\tilde{G}_c] \x \cent[\tilde{G}]_0,
 \edm
 where $\cent[\tilde{G}_c]$ is a finite group which can be determined
 from the root data of $\lalg_c$. The general connected Lie group with
 Lie algebra \lalg\ is then of the form $H=\tilde{G}/D$, where
 $D$  is a discrete subgroup of  $\cent[\tilde{G}]$, and compact if and
 only if $D$ contains a subgroup isomorphic to $\Z^n$.
 A representation of  $\tilde{G}$ gives a
 representation of the factor group $H$ exactly in those
 cases where $D$ operates trivially. Since the elements of $D$ are
 central, they act as multiplication with some scalar on the
 $\tilde{G}$-module under consideration; therefore, $D$ acts trivially
 if and only if this scalar is equal to $1$ for all elements of $D$.
Because of Schur's Lemma, any representation $\pi$ of $\tilde{G}$ can be
described in terms of representations of its center and of its commutator
subgroup. The condition 
that $D$  acts trivially then leads to congruence equations 
between the character of the center and the
size of the representation of the semisimple part.

\subsection{General description of all compact Lie groups of \lalgSM}
\label{core}
%
 Instead of  $\lalgSM$, we may consider the slight generalization
 \bdm
 \lalg (p,q)\ = \ \udef[1] \op \sudef[p] \op \sudef[q],
 \edm 
 where $p$ and $q$ are two different prime numbers. Its universal
 covering group is
 \bdm
 \tilde{G}(p,q) \ \cong\ \R \x \SUdef[p] \x \SUdef[q]
 \text{ with center }
 \cent[\tilde{G}(p,q)] \ \cong\  \R \x \Cycl_p \x \Cycl_q \, .
 \edm
 Then any infinite discrete central subgroup
 $D\leq\cent[\tilde{G}(p,q)]$ leads to a compact factor group
 \begin{equation}\label{G-isom}
 \tilde{G}(p,q)/D=:E(p,q)
 \end{equation}
 with the same Lie algebra as $\lalg (p,q)$. 
 We define $D_1$ to be the intersection of $D$ with the \R-factor of
 $D$, i.\ e.\ $D\cap(\R\x\{e\}\x\{e\})=D_1\x\{e\}\x\{e\}$. It is no loss
 of generality to assume $D_1=\Z$; this means that the homomorphism 
 $\tilde{G}(p,q)\ra E(p,q)$ factors through the map 
  \bdm
  \vphi:\ \tilde{G}(p,q) \lra G(p,q):= \Udef[1]\x\SUdef[p]\x\SUdef[q]
  \edm
 which sends $(\mu, a, b)$ to $(e^{2\pi i \mu},a,b)$. Thus the image
 $\vphi(D)$ of $D$ in $G(p,q)$ satisfies 
 \bdm
 G(p,q)/\vphi(D)\cong E(p,q).
 \edm
 If $z=(e^{2\pi i\mu},\zeta_p^m,\zeta_q^n)\in\vphi(D)$, then 
 $z^{pq}=(e^{2\pi i \mu pq},1,1)=(1,1,1)$ because of our choice of
 $D_1$. Thus $\mu$ must be an integer multiple of $1/pq$, which amounts
 to saying that the group order of $\vphi(D)$ must be a divisor of $pq$.
 This gives us 
 \begin{thm}\label{abstrclass}
 There are exactly nine families of possibilities for $\vphi(D)$, namely:
 \bdm\ba{lclllcll}
 (\mrm{I}) : & 
 \lan (1,1,1)\ran & =: & \mc{I} & & &  \\
 (\mrm{P}_1) : &
 \lan (1,\zeta_p,1)\ran & =: & \mc{P}_1, & 
 (\mrm{P}_2^{(m)}) : &
 \lan (\zeta_p^{-1},\zeta_p^{m},1)\ran & =: & \mc{P}_2^{(m)}\\
 (\mrm{Q}_1) : &
 \lan (1,1,\zeta_q)\ran & =: & \mc{Q}_1, & 
 (\mrm{Q}_2^{(n)}) : &
 \lan (\zeta_q^{-1},1,\zeta_q^{n})\ran & =: & \mc{Q}_2^{(n)}\\
 (\mrm{PQ}_1) : &
 \lan (1,\zeta_p,\zeta_q)\ran & =: & \mc{PQ}_1, & 
 (\mrm{PQ}_2^{(m)}) : &
 \lan (\zeta_p^{-1},\zeta_p^{m},\zeta_q)\ran & =: & \mc{PQ}_2^{(m)}\\
 (\mrm{PQ}_3^{(n)}) : &
 \lan
(\zeta_q^{-1},\zeta_p,\zeta_q^{n})\ran & =: & \mc{PQ}_3^{(n)}, &  
 (\mrm{PQ}_4^{(m,n)}) : &
 \lan (\zeta_p^{-1}\zeta_q^{-1},\zeta_p^{m},\zeta_q^{n})\ran & =: & 
 \mc{PQ}_4^{(m,n)}\\
 \ea\edm
 where the indices $m$ and $n$ can take any of the values
 \bdm
 m\in\mc{M}:=\{1,\ldots,\lbrack (p-1)/2\rbrack\},\quad
 n\in\mc{N}:=\{1,\ldots,\lbrack (q-1)/2\rbrack\}\, .
 \edm
 The respective group orders are
 \bdm
 |\mc{I}| = 1,\ 
 |\mc{P}_1| = |\mc{P}_2^{(m)}| = p,\
 |\mc{Q}_1| = |\mc{Q}_2^{(n)}| = q,\
 |\mc{PQ}_1| = |\mc{PQ}_2^{(m)}| = |\mc{PQ}_3^{(n)}| = |\mc{PQ}_4^{(m,n)}| 
 = pq;
 \edm
 as a group, any nontrivial $\vphi(D)$ is thus isomorphic to either $\Cycl_p$,
 $\Cycl_q$ or $\Cycl_{pq}\cong\Cycl_p\x\Cycl_q$.
\end{thm}
 
 \noindent\textit{Proof.}
 The justification that these are exactly all occuring possibilities for 
 $\vphi(D)$ is elementary for every given group order. We shall therefore 
 not treat them all in detail but just mention that one has to (repeatedly)  
 use the following facts:
 \begin{enumerate}
 \item our choice of $D_1$ rules out any elements of the form 
 $(\zeta_{pq}^l,1,1)\in\vphi(D)$ other than $(1,1,1)$;
 \item complex conjugation is an outer automorphism of $\Udef[1]$ and thus
 does not modify the factor group; we may therefore replace any 
 $\zeta\in\Udef[1]$ by its inverse $\zeta^{-1}$ without changing the factor
 group;
 \item any power of $\zeta_p$ or $\zeta_q$ different from $1$ is a primitive
 $p$th respective $q$th root of unity; 
 \item the Chinese Remainder Theorem implies that for any  $n\in\mc{N}$,
 we can find a coset $a+pq\Z$ such that $a\equiv 1\bmod p$ and
 $a\equiv n \bmod q$; thus, $ \zeta_p^a=\zeta_p$, whereas
 $\zeta_q^a=\zeta_q^n$. Of course, this may be equally applied to a situation
 with $p$ and $q$ exchanged. 
 \end{enumerate}\qed

\noindent
 We thus have the surprising result that all groups $\vphi(D)$ are
 cyclic, which was in general not true for the initial group $D$.

 Now consider any representation of $\lalg(p,q)$, that is, a triple
 consisting of a $\udef[1]$-representation of parameter $\lambda\in\Z$, 
 a $\sudef[p]$-representation of size $r_p$ and a $\sudef[q]$-representation
 of size $r_q$. Recall that these are in one-to-one correspondence with
 representations of $G(p,q)$. Any element $z=(a,b,c)\in\vphi(D)$ will then
 map under this representation to $(a^{\lambda},b^{r_p},c^{r_q})$, and
 because of the isomorphism \eqref{G-isom} and the general remarks in
 Section \ref{allLiegrpsabstract}, this image will operate trivially
 if and only if the product $s=a^{\lambda}b^{r_p}c^{r_q}$ of its factors is
 equal to
 $1$. Since all groups $\vphi(D)$ turned out to be cyclic, it is enough to
 test this condition on the generating elements listed in
 Theorem~\ref{abstrclass}.  We will then get for every possible $\vphi(D)$ a
 necessary and sufficient congruence equation relating the parameters
 $\lambda$, $r_p$, $r_q$, $m$ and $n$. One easily verifies that these are:
%
%
\begin{lem}\label{intcond}
A representation of $\lalg(p,q)$ can be lifted to a representation of 
$G(p,q)/\vphi(D)$ with $\vphi(D)$ of one of the types listed in 
Theorem~\ref{abstrclass} if and only if the corresponding 
congruence equation of the same
type as given below holds:
 \bdm\ba{llll}
 (\mrm{I}) : & 
 - & & \\
 (\mrm{P}_1) : &
 r_p\ \equiv\ 0 \bmod p, &
 (\mrm{P}_2^{(m)}) : &
 m r_p\ \equiv\ \lambda \bmod p, \\
 (\mrm{Q}_1) : &
 r_q\ \equiv\ 0 \bmod q, &
 (\mrm{Q}_2^{(n)}) : &
 n r_q\ \equiv\ \lambda \bmod q, \\
 (\mrm{PQ}_1) : &
 q r_p+p r_q\ \equiv\ 0 \bmod pq, &
 (\mrm{PQ}_2^{(m)}) : &
 q m r_p + p r_q\ \equiv\ q\lambda \bmod pq, \\
 (\mrm{PQ}_3^{(n)}) : &
 q r_p + p n r_q\ \equiv\ p\lambda \bmod pq, &
 (\mrm{PQ}_4^{(m,n)}) : &
 q m r_p + p n r_q \ \equiv\ (p+q)\lambda \bmod pq. \\
 \ea\edm
\qed
\end{lem}
\noindent
 Because these congruence equations yield criteria when a given Lie algebra
 representation can be lifted to a representation of some Lie group, we will
 call them \emph{integrability conditions} (IC).
 Some of the factor groups we can immediately identify with well known
 Lie groups, namely
 \bdm\ba{lclc}
 (\mrm{I}) : & 
 G(p,q) & \cong & \Udef[1]\x\SUdef[p]\x\SUdef[q] \\
 (\mrm{P}_1) : &
 G(p,q)/\mc{P}_1 & \cong & \Udef[1]\x\mrm{P}\SUdef[p]\x\SUdef[q]\\
 (\mrm{P}_2^{(1)}) : &
 G(p,q)/\mc{P}_2^{(1)} & \cong & \Udef[p]\x\SUdef[q]\\
 (\mrm{Q}_1) : &
 G(p,q)/\mc{Q}_1 & \cong & \Udef[1]\x\SUdef[p]\x\mrm{P}\SUdef[q]\\
 (\mrm{Q}_2^{(1)}) : &
 G(p,q)/\mc{Q}_2^{(1)} & \cong & \SUdef[p]\x\Udef[q]\\
 (\mrm{PQ}_1) : &
 G(p,q)/\mc{PQ}_1 & \cong & \Udef[1]\x\mrm{P}\SUdef[p]\x\mrm{P}\SUdef[q]\\
 (\mrm{PQ}_2^{(1)}) : &
 G(p,q)/\mc{PQ}_2^{(1)} & \cong & \Udef[p]\x\mrm{P}\SUdef[q]\\
 (\mrm{PQ}_3^{(1)}) : &
 G(p,q)/\mc{PQ}_3^{(1)} & \cong & \mrm{P}\SUdef[p]\x\Udef[q]\\
 \ea\edm
 where $\mrm{P}\SUdef[n]$ denotes as usual $\SUdef[n]$ modulo its
 center $\mc{Z}\cong\Cycl_n$.
 We will prove later that 
 \bdm\ba{llll}
 (\mrm{PQ}_4^{(1,1)}) : &
 G(p,q)/\mc{PQ}_4^{(1,1)} & \cong & \mrm{S}(\Udef[p]\x\Udef[q]), \\
 \ea\edm
 which can be realized as the elements  $(a,b)\in\Udef[p]\x\Udef[q]$ 
 satisfying
 the additional condition $\det a \det b = 1$, or via a suitably chosen
 Kronecker product. 
\begin{rem}\textit{Interpretation of the parameters $m$ and $n$.}
Every value of $m$ yields a possible  identification of  $\Cycl_p$ in
$\Udef[1]$ with $\Cycl_p$ in $\SUdef[p]$; $m=1$ corresponds to the
identification we are familiar with  by its realization in $\Udef[p]$
(and correspondingly for $n$ and $\Udef[q]$).
The ``twisted'' versions of $\Udef[p]$ we get for $m\neq 1$ are not isomorphic;
they do not appear for the SM, for then we have $\mc{M}=\mc{N}=\{1\}$.
\end{rem}
%
\begin{rem}\textit{Simplification of integrability conditions.}
 Since $p$ and $q$ are coprime, each of the four integrability conditions 
 which are congruence equations $\bmod\,pq$ holds exactly if the 
 same
 relation is true $\bmod\,p$ and $\bmod\,q$ simultaneously; these relations
 in turn may
 be further simplified by eliminating the multiples of $p \bmod p$ or of
 $q \bmod q$ and dividing by any remaining factors coprime to
 $p$ or  $q$, respectively. We get:
 \bdm\ba{llll}
  (\mrm{PQ}_1) : &
 \left\{\ba{l}
  r_p\ \equiv\ 0\bmod p \\r_q\ \equiv\ 0\bmod q\ea\right\} &
 (\mrm{PQ}_2^{(m)}) : &
 \left\{\ba{l}
  mr_p\ \equiv\ \lambda\bmod p \\r_q\ \equiv\ 0\bmod q\ea\right\}\\[5mm]
 (\mrm{PQ}_3^{(n)}) : &
 \left\{\ba{l}
   r_p\ \equiv\ 0 \bmod p \\ n r_q\ \equiv\ \lambda \bmod q\ea\right\} &
 (\mrm{PQ}_4^{(m,n)}) : &
 \left\{\ba{l}
  m r_p\ \equiv\ \lambda \bmod p \\ n r_q\ \equiv\ \lambda \bmod q\ea\right\}
 \ea\edm

\end{rem}

\begin{rem}\textit{Interpretation of integrability conditions.}
 For an \sudef-representation of size $r_n$, the set of points of the weight
 lattice which satisfy the congruence relation $r_n\equiv \lambda \bmod n$ 
 for some integer $\lambda$ is the root lattice in case
 $\lambda\equiv 0\bmod n$ and a translate of it otherwise.
 \end{rem}
%
\begin{rem}\textit{Comparison with results by other authors.}
 Comparing these results in the case $p=2$ and $q=3$ with the
 classification (without proof) in O'Raifeartaigh's book 
 \cite[p.~55 ff.]{O'Raif}, we see that his $G(p,q) / \Cycl_{p+q}$ does
 not appear in the classification above; his further study of the
 subject as well as a footnote in \cite{Hucks91} show that he must have
 meant $G(p,q)/\lan(\zeta_p^{-1}\zeta_q^{-1},\zeta_p,\zeta_q)\ran$ instead.
 \end{rem}
%
\section{Integrability conditions and the Standard Model}
\subsection{The ``minimal'' Lie group of the Standard Model}\label{minLiegrp}
%
 The results above can be directly applied to the Standard Model
 (i.e., $p=2$, $q=3$). The size of an
 $\SUdef[2]_T$-representation is exactly
 $r_T=2t$, that of an $\SUdef[3]_C$-representation $r_C=i+2j$
 (cf. Section~\ref{irredsun}). Every of the nine families of
 compact Lie groups with Lie algebra $\lalgSM$ has only one member,
 which is why we will drop the superscripts $m$ and $n$ from now on.
 A glimpse at Table~\ref{partpropbis}  shows
 that experimentally, conditions ($\mrm{P}_1$) and ($\mrm{Q}_1$) are
 not satisfied, and thus so are all conditons implying them, that is,
 ($\mrm{PQ}_1$), ($\mrm{PQ}_2$) and ($\mrm{PQ}_3$). Ignoring the empty
 condition ($\mrm{I}$), this leaves us with the three possibilities
 ($\mrm{P}_2$), ($\mrm{Q}_2$) and ($\mrm{PQ}_4$), the latter being
 exactly the union of the former two. These conditions do actually hold,
 \begin{equation}\tag{$\mrm{PQ}_4$ a, b}
 \lambda\ \equiv\ r_T \bmod 2\ \text{ and }\ \lambda \equiv\ r_C \bmod 3,
 \end{equation}
 making it possible to define 
 $\grpSM(p,q):=G(p,q)/\lan (\zeta_p^{-1}\zeta_q^{-1},\zeta_p,\zeta_q)\ran$
 and see that 
 \bdm
 \grpSM\ :=\ \grpSM(2,3)\ =\ G(2,3)/\lan (-\zeta_3^{-1},-1,\zeta_3)\ran
 \edm
 is in this sense the ``minimal'' Lie group of the Standard Model, that is,
the smallest Lie group able to explain the experimental evidence. A very 
attractive property of this group is that it establishes hypercharge as the
link between isospin and colour, thus relating quantities which are 
independent in the traditional choice $G(2,3)$. 

We would like to illustrate the logic of our argument by a more familiar
example. To the spin Lie algebra $\sudef[2]$, there correspond exactly two
compact
Lie groups, namely $\SUdef[2]$ and $\mrm{P}\SUdef[2]\cong\SOdef[3]$;
the size of a $\sudef[2]$-representation being twice its spin 
$j\in\frac{1}{2}\N$, the former has empty integrability condition and the
latter has $2j\equiv 0 \bmod 2$, which is equivalent to the condition
that $j$ be an integer. Historically, the experimental evidence of
half integer spins led to the conclusion that the ``right'' Lie group
has to be $\SUdef[2]$ and not $\SOdef[3]$. For the Standard Model, we
experience the reverse situation: assume one would had thought that
$\SUdef[2]$ is the spin Lie group and that experiment had only yielded
integer spin values. Then the logical conclusion would have been that 
$\SOdef[3]$ is a better choice for the acting Lie group than
$\SUdef[2]$ is. In the same vein, we suggest that $\grpSM$ is a better
choice for the inner gauge group of the Standard Model than the usual
choice $G(2,3)$ is. 

From now on, we will \emph{postulate} that $\grpSM$
is the gauge group of the Standard Model and investigate the
conclusions implied by this choice.

 The representation ring of $\grpSM$ has five generators, one possible
 choice for them being
 \bdm
 R(\grpSM)\ \cong\ \Z[V\ox\C_{3},\, W\ox\C_{-2},\, 
 \extp[2]W\ox\C_{-4},\,
 \C_{6},\, \C_{-6}]
 \edm
 where $V$ ($W$) is the $2$- ($3$)-dimensional defining representation of 
 $\SUdef[2]$ ($\SUdef[3]$). They correspond to the winding numbers and
 representation sizes
 \bdm
 (\lambda,r_T,r_C)\ =\ (3,1,0),\ (-2,0,1),\ (-4,0,2),\ (6,0,0),\ (-6,0,0)   
 \edm
 and may be identified with the following lepton and quark fields as
 introduced in Table~\ref{partpropbis}:
 \bdm
 R(\grpSM)\ \cong\ \Z[l(x),\, d(x),\, u^*(x),\, e^*(x),\, e(x)].
 \edm

\begin{rem}\textit{Antiparticles.}
The transition from any particle multiplet to its antiparticle
multiplet is made by taking the dual representation. In our case, this
amounts to replacing the hypercharge by its negative and reversing the
order of the Dynkin indices. From a group theoretical point of view,
it is clear that the dual representation can always be formed. But we may
also deduce the integrability condition for the dual representation by
the following short argument, thus proving that conditions ($\mrm{PQ}_4$)
also hold for the antiparticles which were not listed in 
Table~\ref{partpropbis}: 
assume $\lambda\equiv r_n \bmod\, n$ for a $\SUdef$-representation of
size $r_n$. By taking its negative, we get $-\lambda \equiv
-r_n\bmod\, n$. Now remember that $r_n$ was defined as $\tssum j a_j$;
since of course $-j$ is congruent to $n-j \bmod\, n$, we may rewrite
this as $-\lambda \equiv \tssum (n-j)a_j \bmod\, n$. But then the righthand
side is exactly the size of the representation with reversed
order of Dynkin indices.
\end{rem}

\begin{table}
\bcen
\setlength{\extrarowheight}{5pt}
\setlength{\doublerulesep}{5pt}
\btab{|c|c|c|c|c|c|c|c|}\hline
Quantum field & $y$ & $\lambda=6y$ & $r_T=2t$ & $r_C=i+2j$ & $d_T$ & $d_C$ &
$\lambda d_T d_C$\\
   \hline\hline
l.~h.\ quarks $q(x)$   & $ 1/6$  & $1$  & $1$ & $1$ & $2$ & $3$ & $6$\\\hline
r.~h.\ u-quarks $u(x)$ & $2/3$   & $4$  & $0$ & $1$ & $1$ & $3$ & $12$\\\hline
r.~h.\ d-quarks $d(x)$ & $-1/3$  & $-2$ & $0$ & $1$ & $1$ & $3$ & $-6$\\\hline\hline
l.~h.\ leptons $l(x)$ & $-1/2$ & $-3$ & $1$ & $0$ & $2$ & $1$ & $-6$\\\hline
r.~h.\ leptons $e(x)$ & $ -1$  & $-6$ & $0$ & $0$ & $1$ & $1$ & $-6$\\\hline\hline
Hypercharge $B_{\mu}(x)$ & $0$ & $0$ & $0$ & $0$ & $1$ & $1$ & $0$\\\hline
Isospin $W_{\mu}(x)$     & $0$ & $0$ & $2$ & $0$ & $3$ & $1$ & $0$\\\hline
Colour $G_{\mu}(x)$      & $0$ & $0$ & $0$ & $3$ & $1$ & $8$ & $0$\\\hline\hline
Higgs $h(x)$  & $-1/2$ & $-3$ & $1$ & $0$ & $2$ & $1$ & $-6$\\\hline
\etab

\begin{picture}(250,40)(23,10)
\qbezier(118,48)(138,35)(158,48)
\qbezier(118,40)(164,10)(210,40)
\put(164,39){\makebox(0,0){$\equiv \bmod 2$}}
\put(210,25){\makebox(0,0){$\equiv \bmod 3$}}
\end{picture}
\ecen
\caption{Quantum numbers of the elementary fields, stated in an way appropriate
for representation theory. One then checks easily that the indicated
congruence relations hold.}
\label{partpropbis}
\end{table}

\subsection{Consequences for the electric charge}
%
As an example of the severe restrictions the relations ($\mrm{PQ}_4$)
impose on admissible \grpSM-representations, consider
a bound quark state. As explained in Remark~\ref{quarkconfinrem}, these
can only appear in representations with $r_C \equiv0\bmod\, 3$. 
%
\begin{lem}\label{quarkconfinlem}
For any (usually non irreducible) \grpSM-representation, the condition
$r_C\equiv 0\bmod\, 3$ is equivalent to integer electric charge for all
particles contained in its irreducible subrepresentations.
\end{lem}
\begin{proof}
The statement is an easy  consequence of the relations ($\mrm{PQ}_4$).
We will only show one direction; the
converse  may be proved in a similar way. Remember that $y$ was
defined as $\lambda/6$. Then $r_C\equiv 0\bmod\, 3$ together with 
$r_C\equiv\lambda\bmod\, 3$ implies $2y\in\Z$. Using the Clebsch-Gordan
formula, we see that any irreducible  multiplet of an
$\sudef[2]$-representation of size $r_T$ will have a highest weight $2t$
of same parity. Thus the other congruence relation
$r_T\equiv\lambda\bmod\, 2$ implies $t+3y\in\Z$ for all such $t$,
which together with
$2y\in\Z$ gives $t+y\in\Z$. Since $t_3=t,t-1,\ldots,-t$ is an integer or a
half-integer precisely when $t$ is, we thus have $y+t_3\in\Z$ for all
values of $t_3$.
\end{proof}
 
\begin{rem}\textit{Electric charges.}
 For studying anomalies, an important quantity is the sum of the electric
 charges inside a particle multiplet of given chirality. Unfortunately,
 this number is ill suited for representation theoretical studies,
 since it does not correspond to any characteristic quantity. However, 
 given any $\sudef[2]$-representation with highest weight $2t$,
 the Gell-Mann-Nishijima formula $q=y+t_3$ yields upon summation over
 $t_3=t, t-1,\ldots,-t+1,-t$ the relation $\tssum q = (2t+1)y = d_T y$.
 In order to get the total electric charge for a $\grpSM (p,q)$-multiplet,
 we have to multiply by the dimension of the colour-representation,
 thus obtaining
 \bdm
 \sum_{\grpSM(p,q)\text{ rep.}} q =\ y d_T d_C \, .
 \edm
 The righthand side makes sense for any values of $p$ and $q$ and 
 is accessible to group theoretical arguments as the following theorem
 shows.
\end{rem}

\begin{thm}\label{teilerdim}
 Given a representation of $\grpSM (p,q)$,
 the dimensions of the corresponding representations of $\SUdef[p]$ and
 $\SUdef[q]$ and the winding number satisfy the relation
 \bdm
 pq \ | \ d_p d_q \lambda.
 \edm
\end{thm}
\noindent
By the preceding remark and because $y=\lambda/6=\lambda/pq$, it is clear
that Theorem~\ref{teilerdim} immediately implies for the SM:

\begin{cor}\label{teilerdimSM}

If\ \,$\grpSM$ is the gauge group of the internal SM symmetries, then
the sum over all electric charges inside any of its representations 
(particle multiplets) has to be an integer:
 \bdm
 \sum_{\grpSM(p,q)\text{ rep.}} q\ \in \Z.
 \edm
\end{cor}
\begin{rem}\textit{Covering groups and anomalies.}
An easy consequence of the requirement of anomaly freedom in the SM
is the condition that the sum of $\lambda d_p d_q$ over all
fields of a given chirality has to be exactly zero. By choosing 
$G_{SM}$ instead of $G(2,3)$ as the SM Lie group, Theorem~\ref{teilerdim}
implies that these quantities are congruent $0 \bmod 6$; thus, this choice
does not interfere with the anomaly freedom of the SM.
\end{rem}
\begin{proof}(of Theorem~\ref{teilerdim}) The formula for the dimension of an
 $\SUdef[p]$-representation with Dynkin indices $(a_1,\ldots,a_{n-1})$ is
 \bdm
 d_p \ =  \ \prod_{0\le r < s < p}\left(\sum_{j=r+1}^{s}\tilde{a}_j 
 \right) \cdot \prod_{l=1}^{p}\frac{1}{(l-1)!}
 \edm
 with $\tilde{a}_j=a_j+1$ and its analog for $\SUdef[q]$. We show the
 stronger relations $p\, |\, \lambda d_p$ and $q\, | \, \lambda d_q$,
 which imply the assertion since $p$ and $q$ are two different
 prime numbers. Because of the symmetry of the problem, it is enough 
 to show the assertion for $p$. Relation ($\mrm{PQ}_4$ a) implies
 that the difference of $\lambda$ and $r_p$ is a multiple of $p$ and
 it is therefore enough to show $p \, | \, r_p d_p$.
 The factorials in the denominator of $d_p$ contain only
 factors $< p$ and can thus be ignored, leaving us with the claim
 \begin{equation}\label{p-teiler}
 p \ \text{ divides } \ r_p \quad \cdot \!\!\!\prod_{0\le r < s\le p-1}\left(
 \sum_{j=r+1}^{s}\tilde{a}_j \right).
 \end{equation}

\noindent
Case $p=2$ : Equation~\eqref{p-teiler} is reduced to $2 \ | \ r (r+1)$,
$r$ the size of the representation, and this is of course always true.
 
\noindent
Case $p\neq 2$ : Now we have
 \bdm
 \sum_{1}^{p-1} j \tilde{a}_j \ =\ \sum_{1}^{p-1} j (a_j+1) \ =\ 
 \sum_{1}^{p-1} j a_j + \sum_{1}^{p-1} j \ =\ r_p + \frac{p(p-1)}{2}.
 \edm
 Since $p$ is an odd prime number, $2\ | \ p-1$. Therefore $p$
 divides the last term and it disappears $\bmod\, p$:
 \bdm
 \sum_{1}^{p-1} j \tilde{a}_j \ \equiv\ r_p \bmod p .
 \edm
 Rewrite the negative of this last expression as
 \bdm
 - \sum_{1}^{p-1} j \tilde{a}_j \ 
 \equiv\  \sum_{1}^{p-1} (p-j) \tilde{a}_j \
 \equiv\ \left\{\ba{l}  \phantom{+(}0 + \\
 +(0+\tilde{a}_1)+ \\
 +(0+\tilde{a}_1+\tilde{a}_2)+\\
 +\ldots +\\
 +(0+\tilde{a}_1+\ldots+\tilde{a}_{p-1})\, . \ea\right.
 \edm
 If all $p$ terms on the righthand side are different modulo $p$, 
 their sum is congruent to $\frac{p(p-1)}{2} \bmod p$ and therefore
 equal to $0 \bmod p$ by the argument above, and we have thus
$r_p  \equiv 0 \bmod p$. If not, there exist indices $0\le r<s<p$ satisfying
 \bdm
 0+\tilde{a}_1+\ldots+\tilde{a}_r = 0+\tilde{a}_1+\ldots+\tilde{a}_s \bmod p,
 \text{ thus }
 \sum_{j=r+1}^{s} \tilde{a}_j \ \equiv\ 0 \bmod p,
 \edm
 and $d_p$ is divisible by $p$.
\end{proof}
 
 \bigskip\noindent
 Although only based on simple number theoretical steps, this proof
has the disadvantage of not showing any deeper structure. We will
now give two explicit realizations of the group \grpSM. The first
one, very popular, is based on a direct sum construction and extensively
used in connection with ``Grand Unified Theories''. The second one,
less known, will make use of a Kronecker product and give a nice
representation theoretical proof of Theorem~\ref{teilerdim}. 

\section{Realizations of $\grpSM(p,q)$ and Second Proof of 
Theorem \ref{teilerdim}}
\subsection{Additive realization of $G(p,q)$ and its quotient groups}
\label{addreal}
Embed $\SUdef[p]$ as upper left, $\SUdef[q]$ as lower right block
in $\SUdef[p+q]$ and write such elements succintly as pairs
$(a,b)$. For realizing  \grpSM\ as a subgroup of $\Udef[p+q]$,
the requirement that
 $\zeta_p$ should be mapped to $(\zeta_p^{1}\cdot 1_p, 1_q)$ 
 and $\zeta_q$ to $(1_p,\zeta_q^{1}\cdot 1_q)$
 can only be achieved by a map of the form
 \bdm
 e^{it}\lmapsto (e^{ikt}1_p,e^{ilt}\cdot 1_q)
 \quad k,\,l\in\Z
 \edm
 where $k$ and $l$ satisfy $ k\in q\Z \cap ( 1+p\Z)$ and 
$l\in p\Z \cap (1+q\Z)$.
 According to the Chinese Remainder Theorem, there always exist cosets
 $k +pq\Z$ and $l+pq\Z$ fulfilling these conditions.
 Since $p+q$ is coprime to $p$ and $q$, we can further assume 
 $pk+qk\ \equiv\ 0\bmod (p+q)$ and still get solutions which will now be
 cosets $\bmod\, p q (p+q)$, thus making it possible to impose
 trace zero condition upon the $\udef[1]$-generator
 (for $p=2$ and $q=3$, a  possible choice is
 $k=3$ and $l=-2$). This achieves to show the isomorphism between
$\grpSM (p,q)$ and $\mathrm{S}(\Udef[p] \x \Udef[q])$.

\subsection{Multiplicative realization of $\grpSM$}\label{multreal}
We now discuss a realization of \grpSM\ using the Kronecker product
of matrix groups (tensor product of the underlying vector spaces).

%
 
 For $g\in\SUdef[p],\, h\in \SUdef[q]$, let $g\ox h$ act on the 
 tensor product $V\ox W$ of their standard modules. Then the image of
 $(\zeta_p,\zeta_q)$ is
 \bdm
 (\zeta_p\cdot 1_p)\ox (\zeta_q\cdot 1_q) \ = \ \zeta^{p+q}_{pq}\cdot 1_{pq}.
 \edm
By defining the action of $\Udef[1]$ on $V\ox W$ as scalar multiplication, we
get a natural identification of
 $\Cycl_p\x \Cycl_q\subset \SUdef[p]\x \SUdef[q]$ with
 $\Cycl_{pq}\subset\SUdef[pq]$, because $\zeta^{p+q}_{pq}$ is always a
 primitive $pq$th root of unity. To check that it really satisfies the
 integrability conditions ($\mrm{PQ}_4^{(1,1)}$), we first notice that for
 representations with sizes $r_p$, $r_q$ and winding number $\lambda$, the
 following diagram has to commute:
 \bdm
 \begin{CD}
 (\zeta_p,\,\zeta_q)@>>>\zeta_{pq}^{p+q}\\
 @V{r_p,\,r_q}VV @VV{\lambda}V\\
 (\zeta_p^{r_p},\,\zeta_q^{r_q})@>{!}>>\zeta_{pq}^{\lambda(p+q)}
 \end{CD}
 \edm
 which is equivalent to the requirement that the mapping relation
 denoted by ! holds. But
 \bdm
 (\zeta_p^{r_p},\,\zeta_q^{r_q})\ \lmapsto\ 
 \zeta_p^{r_p}1_p\ox \zeta_q^{r_q}1_q 
 \ = \
 \zeta_{pq}^{q r_p + p r_q}1_{pq} 
 \ \stackrel{!}{=} \ 
 \zeta_{pq}^{\lambda(p+q)}1_{pq},
 \edm
 means exactly
 \begin{equation}\label{pq-intequiv}
 q r_p + p r_q \ \stackrel{!}{\equiv}\ \lambda (p+q) \bmod pq\, ,
 \end{equation}
 which we recognize to be the integrability condition ($\mrm{PQ}_4^{(1,1)}$).
 Thus, the Kronecker product of $\SUdef[p]$ and $\SUdef[p]$ with
 a natural action of $\Udef[1]$ by scalar multiplication is isomorphic
 to $\grpSM (p,q)$.
 
\begin{rem}\label{secondproof}
\textit{Second proof of Theorem~\ref{teilerdim}}.
We may reprove Theorem~\ref{teilerdim} with a
purely representation theoretical argument, without even having to
know the formula for the dimensions of the representation.

Indeed, if we have $pq \mid d_p d_q$, then there is nothing to prove. If not,
 assume for example that $p$ does not divide $d_p$. The image of $\zeta_p$
 under an $\SUdef[p]$-representation is the matrix $\zeta_p^{r_p}\cdot
 1_{d_p}$; its determinant has to be $1$, and since $p$ did not divide
 $d_p$, this can only be the case if $\zeta_p^{r_p}=1$. But then the
 images of all $p$th roots of unity have to act trivially under any
 $\lambda$-representation, which amounts to saying that $\lambda$ is a
 multiple of $p$. The same argument holds for $q$.
\end{rem}

\subsection{Non-standard decompositions of compact connected groups.}
\label{zerlegkompgrp}
We are interested in finding Cartan subgroups $H$ of any compact connected
reductive Lie group $G$ which are a direct product of their
intersections with the commutator subgroup and the center. Certainly, 
every complement of the commutator subgroup yields such a Cartan subgroup:
%
\begin{lem}[Direct decompositions of Cartan subgroups]
 Let $G$ be a compact connected reductive Lie group and \kom\ a complement
of its commutator subgroup $G_c$, i.\ e., we have a semidirect decomposition
$G \cong G_c \rx \kom$. Then any Cartan subgroup $H$ of $G$ is the direct
product of its intersection $H_c$ with $G_c$ and \kom, i.\ e.,
$H \cong H_c \x \kom,\  H_c = G_c \cap H$.
\qed
\end{lem} 
%
\subsection{Application to $G=\mathrm{U}(n)$ and electroweak interactions.}
For the Standard Model, it will turn out that these purely group
theoretical considerations have the property of singling out exactly
those symmetries which remain after spontaneous symmetry breakdown 
and Higgs mechanism. We take the liberty of assuming colour
confinement, that is, $\grpSM$ is reduced to the subgroup $G=\Udef[2]$
of the electroweak forces.  Since we thus only need to find
complements of the commutator subgroup $G_c=\SUdef$ of $G=\Udef$,
we refer the reader to the general results by K.~H.~Hofmann and H.~Scheerer 
\cite[Kor.~8]{Hofmann75}, \cite[Lemma]{Scheerer73} without stating them here
for the verification that the following construction yields indeed all
desired complements.

 For a maximal torus $H_c$ of \SUdef\ we make the usual choice of the 
 diagonal matrices in \SUdef. Furthermore, the center $\cento$ of $G$ is
 \Udef[1] and connected, thus giving
 $\cento\cap H_c = \cent[\SUdef]\cong \Cycl_n$. The construction of all
 complements uses a continuous group morphism $f:\ \cento \ra H_c$
 \bdm
 e^{i\omega}\cdot 1_n \stackrel{f}{\lmapsto} \left(\ba{ccc}
 e^{ik_1\omega} & & 0 \\
  & \ddots & \\ 0& & e^{ik_n\omega} \ea \right) \ , \ k_l\in\Z \ 
  \qqs l=1,\ldots,n
 \edm
 satisfying the conditions that the determinant be equal to $1$
 \begin{equation}
 \sum_{l=1}^n k_l= 0 
 \label{firstcond}
 \end{equation}
 and that $f$ be equal to the identity map on $\cent[\SUdef]$ 
 \begin{equation}
 k_1,\ldots,k_n \ \equiv \ 1 \bmod n.
 \label{secondcond}
 \end{equation}
 We then have the representatives
 \bdm
 \{f(z)^{-1}z : z\in\cento\} \ = \ \left\{\left(\ba{ccc} 
 e^{i(1-k_1)\omega} & & 0 \\
  & \ddots &  \\
 0& & e^{i(1-k_n)\omega}\ea\right)\ :\ \omega\in [0,2\pi[\;\right\}
 \edm
 of a class of complements of $G_c$. By substituting 
 $1-k_i=z_i n,\,z_i\in\Z$, condition~\eqref{secondcond} holds
 automatically and since we then have $\tssum k_i = n(1-\tssum z_i)$,
 eq.~\eqref{firstcond} implies $\tssum z_i=1$:
\bdm
 \kom(z_1,\ldots,z_n) \ := \ 
 \{f(z)^{-1}z : z\in\cento\} \ = \ \left\{\left(\ba{ccc} 
 e^{iz_1 n\omega} & & 0 \\
  & \ddots &  \\
 0& & e^{iz_n n\omega}\ea\right) =: k(\omega)\ :\ \ba{c}\omega\in
 [0,2\pi[ \\[3pt]
 \tssum z_i = 1  \ea\right\}
 \edm
 The solutions for $(z_1,\ldots,z_n)$ are, up to permutations,
 \bdm
 (z_1,\ldots,z_n)\ =\ 
 (1,0,\ldots,0), \
 (1-z,z,0,\ldots,0),\ 
 (1-z-z',z,z',0,\ldots,0) \ldots
 \edm
 and each of them defines a complement of \SUdef\ isomorphic to $\Udef[1]$ 
 with the product
 \begin{equation}
 (g,k)(g',k')\lmapsto (g\cdot kg'k^{-1},kk') \ \ \ \qqs g, g' \in \SUdef,\,
 k,k' \in \kom(z_1,\ldots,z_n) \ .
 \label{semicomp}
 \end{equation}
 Because of $k^{-1}(\omega)=k(-\omega)$, we have for 
 $g'=(g'_{ij})_{ij},\, 1\le i,j \le n$:
 \bdm
 kg'k^{-1} \ = \ \left( e^{i\omega n(z_i-z_j)}g'_{ij} \right)_{ij} \ ,
 \edm
 and $kg'k^{-1}$ is certainly equal to $g'$ if $g'$ is diagonal. 
 In this case the composition law \eqref{semicomp} becomes
 \bdm
 (g,k)(g',k')\lmapsto (gg',kk') \ ,
 \edm
 which leads to the following direct product in \Udef:
 \bdm
 H_c \x \kom(z_1,\ldots,z_n) \ < \ \SUdef \ \rx \ \kom(z_1,\ldots,z_n) 
 \ = \ \Udef \,.
 \edm
 Since $H_c \x \kom$ is an $n$-parameter abelian subgroup of $G$, it
is necessarily a maximal torus.

\medskip\noindent
 For $n=2$, the Cartan subgroups of $\SUdef[2]$ are
 one-dimensional and the complements depend only on one parameter $z\in\Z$:
 \bea[*]
 H_c \ = \ \ \left\{\ \left(
 \ba{cc} e^{i\omega} & 0 \\ 0 & e^{-i\omega}\ea\right) \ \right\}, & 
 \lalg[H] \ = \left\{ \omega h, \ \ h=i\left(
 \ba{cc} 1 & 0 \\ 0 & -1 \ea\right) \ \right\} \\
 \kom_z \ = \ \left\{\left(
 \ba{cc} e^{i2z\alpha} & 0 \\ 0 & e^{i2(1-z)\alpha}\ea\right)\right\}, &
 \lalg[K]_z \ = \left\{ \alpha k_z, \ k_z=i\left(
 \ba{cc} 2z & 0 \\ 0 & 2(1-z) \ea\right) \right\} 
 \eea[*]
 Physically speaking, $h$ is the generator of the third component of
isospin and thus has eigenvalue $t_3$. Hypercharge is identified with
the center of $\Udef[2]$. But then the relation
 \begin{equation}
 1_2 + h \ = \ k_1 \ \Lra \ y + t_3 \ = \ \kappa_1 \ ,
 \label{semidirectgen}\end{equation}
 where $\kappa_1$ is the eigenvalue of $k_1$, allows us to identify 
 $\kappa_1$ according to the Gell-Mann-Nishijima-formula with the
 electromagnetic charge $q$:
 \bdm
 H_c \x \kom_1 < \SUdef[2] \rx \kom_1 \ \Lra \ 
 \Udef[1]_T \x \Udef[1]_Q < \SUdef[2] \rx \Udef[1]_Q .
 \edm
 This decomposition may be a hint \emph{why} the transition from 
 hypercharge to electric charge is necessary in the Standard Model.
 It has the property that any element of $\Udef[2]$ may be uniquely
 written as an element of $\SUdef[2] \rx \Udef[1]_Q$, whereas this was true
 only up to central elements for $\SUdef[2] \cdot \Udef[1]_Y$. Of course,
 we cannot explain the breakdown of isospin symmetry in this way, since
 we left the semisimple part untouched. Conversely, we can observe
 empirically: the Higgs mechanism singles out exactly the complement
 of the semidirect part in these special decompositions and forgets
 the semisimple part afterwards.

\noindent
\textit{Acknowledgments.} 
Some of the
physical ideas in this paper are partly due to Heinrich Saller
\cite{Saller92b}, \cite{Saller93b}, to whom I am indebted for many
discussions on the subject. I also thank Rutgers University and especially 
Roe Goodman for his kind hospitality during the writing of this paper,
Friedrich Knop for finding an error,
as well as Karl-Hermann Neeb (Universit\"at Erlangen-N\"urnberg) for
his continued support.


%
\end{document}